\documentclass[twocolumn,showpacs,preprintnumbers,amsmath,amssymb]{revtex4}

\usepackage{graphicx}
\usepackage{dcolumn}
\usepackage{bm}

\begin{document}

\title{All-optical runaway evaporation to Bose-Einstein condensation}
\author{J.-F. Cl\'ement}
\author{J.-P. Brantut}
\author{M. Robert de Saint Vincent}
\author{R.A. Nyman}
\altaffiliation[present address : ]{Center for Cold Matter, Imperial College, London, SW7 26W, UK}
\author{A. Aspect}
\author{T. Bourdel}
\author{P. Bouyer}
\affiliation{
Laboratoire Charles Fabry de l'Institut d'Optique, Univ Paris Sud, CNRS, campus polytechnique RD128 91127 Palaiseau France
}

\date{\today}

\begin{abstract}
We demonstrate runaway evaporative cooling directly with a tightly
confining optical dipole trap and achieve fast production of
condensates of 1.5$\times 10^5$ $^{87}$Rb atoms. Our scheme is
characterized by an independent control of the optical trap
confinement and depth, permitting forced evaporative cooling
without reducing the trap stiffness. Although our configuration is
particularly well suited to the case of $^{87}$Rb atoms in a
1565\,nm optical trap, where an efficient initial loading is
possible, our scheme is general and should allow all-optical
evaporative cooling at constant stiffness for most species.
\end{abstract}


\maketitle

Far-Off Resonance optical dipole Traps (FORT) are used extensively
in ultra-cold atom experiments. They allow great versatility of
the trapping potentials and therefore offer the possibility to
study numerous physical situations such as double wells
\cite{Esteve08}, 2D traps \cite{Hadzibabic06} or artificial
crystals of light \cite{Greiner02}. Moreover, they have the
advantage over magnetic traps that they leave the magnetic field
as a degree of freedom. As a consequence, they are crucial to the
study of Bose-Einstein condensates (BEC) with internal spin
degrees of freedom \cite{Stenger98} or formation of ultra-cold
molecules by means of magnetically-tuned Feshbach resonances
\cite{Donley02, Bourdel03, Regal03}.

All-optical cooling methods to achieve quantum degeneracy have
been successfully implemented in ultra-cold atom experiments, both
for bosonic \cite{Barrett01} and fermionic species
\cite{Granade02}. These methods have several advantages. The
absence of a magnetic trap permits a better optical access to the
trapped cloud and a better control of residual magnetic field for
precision measurements. The optical trap tight confinement allows
fast evaporation ramps and therefore cycling time of only a few
seconds. In addition, for some atomic species, condensation in a
magnetic trap is not possible and all-optical trapping and cooling
is necessary. This is the case for magnetically untrappable
spinless atoms such as Ytterbium \cite{Takasu03} or for Cesium
because of a large inelastic collision rate \cite{Weber03}.

With all these advantages, all-optical evaporative cooling is
severely hindered by the fact that the trap confinement is reduced
during forced evaporative cooling. This is because the reduction
of the trap depth is usually performed by lowering the trap power
(typically by several orders of magnitude), which consequently
reduces the trap frequencies, the collision rate, and therefore
also the evaporation efficiency. This difficulty was overcome in
early experiments with an efficient optical trap loading leading
to both a high phase-space density and a high collision rate
\cite{Barrett01}. Many strategies have been used to improve the
number of condensed atoms\,: Raman sideband cooling in a 3D
lattice to lower the temperature before the transfer the optical
trap \cite{Kinoshita05}, a mobile lens to change the optical trap
waist dynamically \cite{Kinoshita05, Chang06}, or the use of a
second dipole trap to compress the cloud after some evaporation
\cite{Weber03}.

\begin{figure}[hbtp!]
\centering
\includegraphics[width=0.49\textwidth]{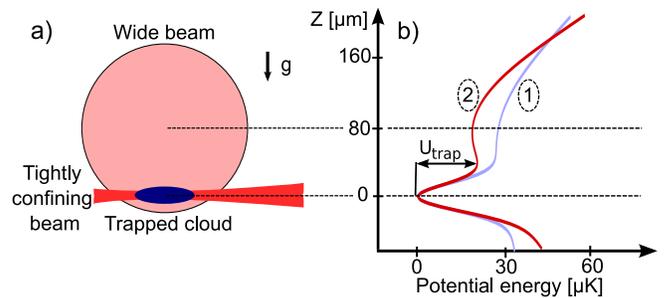}
\caption{a: Scheme of our optical dipole trap. The confining beam
crosses the wide beam at a distance of about 80\,$\mu$m from its
center. This configuration decouples the control of the trap depth
$U_{\textrm{trap}}$ from the control of the trap confinement. b:
Vertical potential energy cuts. The two curves correspond to the
same power of 0.15\,W in the confining beam whereas the wide beam
powers are 8\,W (light/blue curve 1) and 16\,W (dark/red curve 2).
It shows that the trap depth is reduced by increasing the wide
beam power. The atoms, pulled upward from the confining beam
waist, are lost in the direction of the wide beam, perpendicular
to the figure, along which there is no confinement.}
    \label{fig:FORT}
\end{figure}
In this paper, we present a method to decouple the control of the
trap depth from the control of the trap confinement, in analogy to
the case of radio-frequency evaporation in a magnetic trap. We are
thus able to reach the runaway regime, where the collision rate
increases during the evaporation, a situation which seems
impossible to achieve in a single-beam optical trap
\cite{OHara01}. Our all-optical evaporation procedure does not
rely on an external magnetic field gradient in contrast to
\cite{Hung08}. It allows spin-mixture cooling, is fully compatible
with the high trapping frequencies used in most all-optical
cooling experiments and can be straightforwardly generalized to
all-optical cooling of most atomic species.

Our trap is composed of two horizontal crossing beams, a wide beam
and a tightly confining beam, with waists of 180\,$\mu$m and
26\,$\mu$m respectively (fig.\,\ref{fig:FORT}a). The tightly
confining beam is offset by $\sim$80\,$\mu$m from the center of
the wide beam as sketched in fig.\,\ref{fig:FORT}. It is
responsible for most of the trap confinement and the atoms are
trapped at its waist. On the contrary, the wide beam is used to
control the trap depth $U_{\textrm{trap}}$. It applies locally a
force, which pulls the atoms out of the confining beam. By varying
its power we can vary the trap depth and it acts as an effective
evaporative knife. During the evaporation, we can thus control the
trap depth and confinement independently. This method allows us to
implement an efficient forced evaporation in the runaway regime to
Bose-Einstein condensation.

In addition, our peculiar optical trap geometry is well suited to
an efficient loading from the magneto-optical trap. This is done
in two steps. We first combine the wide optical beam with an
extremely far detuned optical molasses. It takes advantage of a
position selective optical pumping due to the strong light-shifts
induced by the wide trapping beam at 1565\,nm \cite{Brantut08}.
Second, the atoms are transfered into the tightly confining beam
creating the trap configuration used for the evaporation.


\begin{figure}[t!]
\centering
\includegraphics[width=0.45\textwidth]{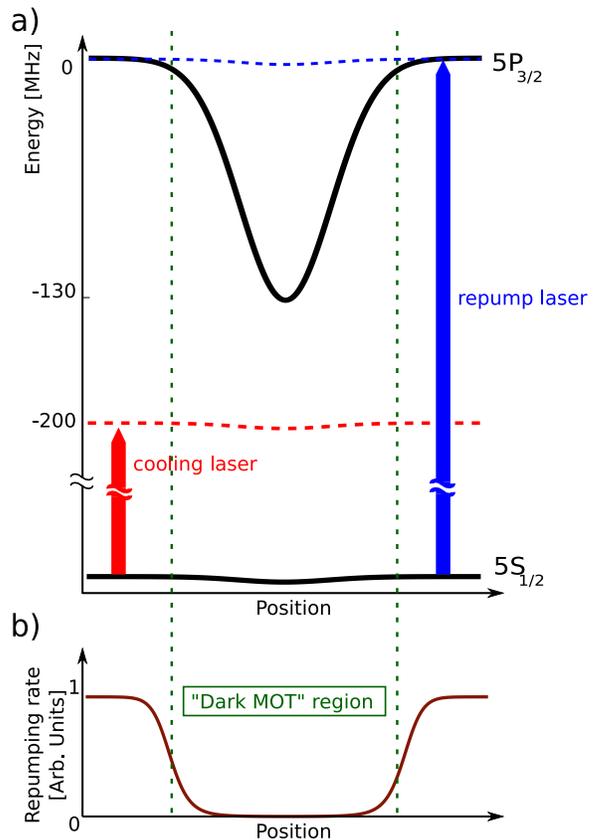}
\caption{Loading scheme into the wide trapping beam. a:\,\,Light
shifted levels of the 5S$_{1/2}$ and 5P$_{3/2}$ states of
$^{87}$Rb under the influence of the wide beam, at a power of
28\,W. At 1565\,nm, the 5P$_{3/2}$ excited state of the MOT
cooling transition is 42.6 more shifted than the 5S$_{1/2}$
fundamental state \cite{Brantut08}. For clarity the hyperfine
structure is not shown. During the very far detuned molasses
phase, the cooling laser is 200\,MHz detuned and therefore remains
to the red of the cycling transition even in the presence of the
trapping laser. The repumping light on the contrary is on
resonance in free space and is brought out of resonance by the
dipole laser light. b: Repumping rate as a function of position.
The repumping is not efficient at the position of the dipole trap
leading to a depumping effect leaving the atom into the ``dark''
$F=1$ hyperfine states. }
    \label{fig:light_shift_approach}
\end{figure}

%
The experimental sequence begins with a standard 3D-MOT ($\sim
3\times 10^9$ atoms), loaded from a 2D-MOT in a few seconds, as
described in \cite{Brantut08}. The detuning of the cooling beams
is then increased to 120\,MHz for 60\,ms, which leads to a
compression of the cloud as the repelling forces due to multiple
scattering are reduced \cite{Petrich94}. At this point starts our
dipole trap loading procedure. The wide trapping beam at 1565\,nm
with a waist of 180\,$\mu$m and a power of 28\,W is turned on.
Simultaneously the magnetic field gradient is turned off and the
cooling laser detuning is increased to 200\,MHz. This value is
chosen such that the cooling beam remains red detuned even in the
presence of the strong light-shift induced by the trapping laser
(Fig.\,\ref{fig:light_shift_approach}a). This corresponds to a
very far detuned optical molasses with the dipole trap beam
intersecting the atomic cloud in its center. It lasts for 50\,ms.

Our dipole trapping beam not only affects the cooling transition
but also detunes the repumping beam out of resonance, as shown in
Fig.\,\ref{fig:light_shift_approach}a. The repumping efficiency
decreases as the FORT laser intensity increases
(Fig.\,\ref{fig:light_shift_approach}b) and the atoms are pumped
to the $F=1$ hyperfine states. This reduces scattered photon
reabsorption, a process which limits the density of laser-cooled
samples. We thus create an effective spatial dark MOT
\cite{Ketterle93} induced by the trapping laser itself
\cite{Barrett01, Couvert08}. The depumping effect is further
enhanced by reducing the repumping laser intensity by a factor
$\sim30$ to 21\,$\mu$W.cm$^{-2}$. In absorption imaging without
repumping, we find that 99\% of the atoms remaining in the dipole
trap are in the 5S$_{1/2}$ (F=1) hyperfine states. For the
untrapped atoms this number is only 97\% showing that the dipole
trapping laser causes an additional reduction of the repumping
rate.

The cooling and repumping beams are then switched off and about
$3\times10^7$ atoms remain in the wide trapping beam. Their
temperature is 20\,$\mu$K and the phase space density is $\sim
2\times 10^{-5}$. Longitudinally they occupy a region of about
1\,mm in size, but the trapping force in this direction is
negligible and the atoms can escape along the beam. In order to
create a trap in three dimensions, we add the tightly confining
beam immediately after switching off the molasses. It intersects
the wide beam at an angle of 56$^\circ$. We choose a quite small
waist of 26\,$\mu$m and a power of 6\,W \cite{cross_loading} in
order to create a tight trap, whose frequencies are 124\,Hz,
3.8\,kHz and 3.8\,kHz. After a waiting time of 100\,ms, we end up
with $3\times 10^6$ atoms at 65\,$\mu$K whereas the rest of the
atoms are lost. The phase-space density is $2.5\times 10^{-3}$.
Adding the tightly confining beam has thus allowed us to increase
the phase space density by two orders of magnitude, with only a
factor of 10 reduction in atom number. We attribute this to an
efficient free-evaporative cooling.

This result is non-intuitive as the initial collision rate in the
wide beam is too low to play any role in the loading process and
atoms which fall into the tightly confining beam have enough
energy to escape from it. However, a 1-body calculation using our
trapping potential showed that these atoms follow non trivial 3D
trajectories and typically remain in the crossed-trap region for a
time much longer than the 100\,ms loading time. As a consequence,
the atomic density and the collision rate quickly increases in the
tightly confining trap. The atoms then thermalize and cool down by
evaporation. The final collision rate is greater than
10$^4$\,s$^{-1}$ and the conditions are very favorable for
efficient evaporative cooling.


\begin{figure}[htbp!]
\centering
\includegraphics[width=0.40\textwidth]{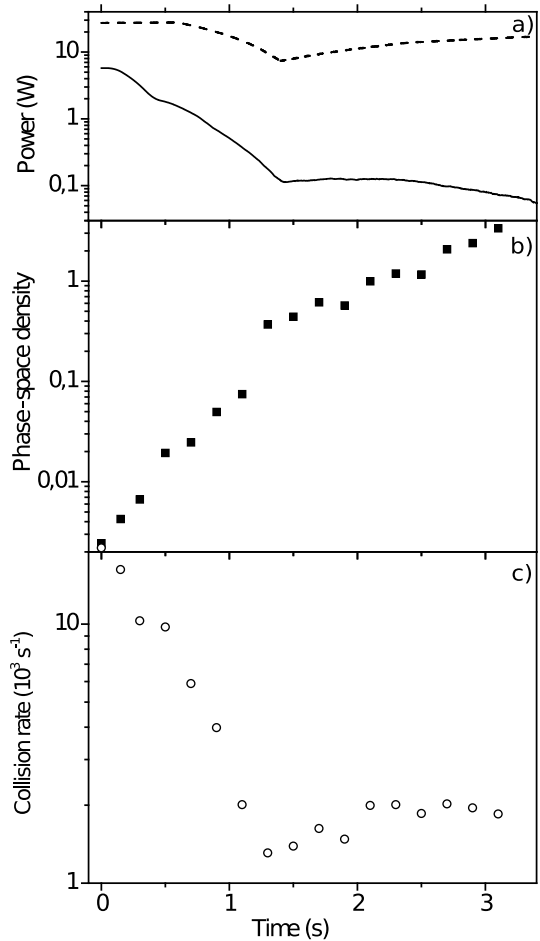}
\caption{a: Optical power in the two trapping beam as a function
of time during the evaporation. The solid (dashed) curve
corresponds to the tightly confining (wide) beam. In the second
part of the evaporation the tightly confining beam power is
roughly constant whereas the power of the wide beam is increased
to force the evaporation. b and c: Phase-space density and
collision rate as a function of time.}
    \label{fig:figure_puissance}
\end{figure}

We now proceed to the final forced evaporative cooling stage. It
takes advantage of our specific trap geometry, which permits
independent control of trap stiffness and depth. The collision
rate and the space-density during the evaporation are presented in
Fig.\,\ref{fig:figure_puissance}. The time sequence is the result
of an optimization of the number of atoms obtained in the
condensate. The evaporation can be decomposed in two phases. In a
first step (which lasts 1.4\,s), we decrease the power of both
beams. Evaporative cooling is then accompanied with a reduction of
the confinement, of the density, and of the collision rate. Such a
reduction is useful as it avoids 3-body losses \cite{power}. In a
second step (which lasts 2\,s) we increase the power of the wide
beam while the tightly confining beam is approximately kept at a
constant power. As explained in the introduction, because of our
original trap geometry, this procedure yields forced evaporation
at constant confinement (Fig.\,\ref{fig:FORT}).

Runaway evaporation is experimentally observed as the collision
rate increases during this phase before it saturates due to 3-body
losses (Fig.\,\ref{fig:figure_puissance}b). Our evaporation ramp,
which lasts 3\,s, leads to $3\times 10^5$ atoms at the critical
temperature and to pure condensates of $1.5\times 10^5$ atoms.
However, by simply reducing the duration of all ramps we were able
to achieve Bose-Einstein condensation in an evaporation as short
as 650\,ms. Our condensates are spinor condensates with relative
abundance of 0.45,0.35, and 0.2 in the $m_F=-1$, $m_F=0$, and
$m_F=1$ magnetic states of the $F=1$ manifold. In order to produce
a BEC in a single state, a spin distillation method
\cite{Couvert08} could be used.

We characterize the efficiency of our evaporation ramp through the
scaling parameter $\gamma = - \frac{\textrm{d} \, \textrm{Ln}
D}{\textrm{d}\,\textrm{Ln} N} = 2.8(5)$ where D is the phase-space
density and N the number of atoms. Given our estimated truncation
parameter $\eta = 11(2)$ (ratio of trap depth to temperature), the
efficiency is lower than expected according to the scaling laws
\cite{OHara01}. The reduction can be understood by including
three-body losses in the analysis. Using a similar treatment to
\cite{Walraven} and \cite{Ketterle96}, the efficiency of the
evaporation is given by
\begin{equation}
\gamma = \eta - 4 - R\,(\eta-2),
\end{equation}
The factor $\eta-4$ comes from the evaporation \cite{OHara01}
while the second term is due to losses; $R$ is the ratio of three
body losses to the total number of atoms removed from the trap
\cite{footnote3body}. With $\eta = 11(2)$, we recover our observed
efficiency of the evaporation with $R\approx0.45$. Such a number
is compatible with our calculated 3-body loss rate of about
1\,$s^{-1}$ at the end of the evaporation.

In conclusion, we have demonstrated an efficient and simple
all-optical route to Bose-Einstein condensation of $^{87}$Rb in a
1565\,nm dipole trap \cite{1565}. A very far-off resonance optical
molasses, taking advantage of the strong light-shifts, is used to
efficiently load the optical dipole trap. Then we apply a new
off-centered crossed-beam configuration, which permits all-optical
runaway evaporation. This method can be straightforwardly
generalized to numerous situations, such as the cooling of
paramagnetic atoms, spin mixtures, atomic mixtures or molecules.
The ability to achieve high duty cycle without a magnetic trap
opens the perspective to use BECs in high precision measurement
applications such as atomic clocks or accelerometers
\cite{LeCoq06}.

We acknowledge G. Varoquaux for his help in the construction of
the experimental apparatus, F. Moron, A. Villing for technical
assistance, G. Lucas-Leclin, D. Gu\'ery-Odelin, and J. V. Porto
for helpful discussions, ImagineOptics company for lending us a
HASO-NIR wavefront analyzer. This research was supported by CNRS,
CNES as part of the ICE project, the project "blanc"
M\'elaBoF\'erIA from ANR, IFRAF, QUDEDIS; by the STREP program
FINAQS of the European Union and by the MAP program SAI of the
European Space Agency (ESA).


\begin{thebibliography}{}

\bibitem{Esteve08} J. Esteve, C. Gross, A. Weller, S. Giovanazzi, and M.K. Oberthaler, Nature \textbf{455}, 1216 (2008)

\bibitem{Hadzibabic06}  Z. Hadzibabic, P. Kr\"uger, M. Cheneau, B. Battelier, and J. Dalibard, Nature \textbf{441}, 1118 (2006)

\bibitem{Greiner02} M. Greiner, O. Mandel, T. Esslinger, T.W. H\"ansch, and I. Bloch, Nature \textbf{415}, 39 (2002)

\bibitem{Stenger98} J. Stenger, S. Inouye, D.M. Stamper-Kurn, H.-J. Miesner, A.P. Chikkatur, and W. Ketterle, Nature \textbf{396}, 345 (1998)

\bibitem{Donley02}E.A. Donley, N. R. Claussen, S.T. Thompson, C.E. Wieman, Nature \textbf{417}, 529 (2002)

\bibitem{Bourdel03} T. Bourdel, J. Cubizolles, L. Khaykovich, K.M.F. Magalhaes, S.J.J.M.F. Kokkelmans, G.V. Shlyapnikov, C. Salomon, Phys. Rev. Lett. \textbf{91}, 020402 (2003)

\bibitem{Regal03} C. A. Regal, C. Ticknor, J.L. Bohn and D.S. Jin, Nature \textbf{424}, 47 (2003)

\bibitem{Barrett01} M.D. Barrett, J.A. Sauer, and M.S. Chapman, Phys. Rev. Lett. \textbf{87}, 010404 (2001)

\bibitem{Granade02} S.R. Granade, M. E. Gehm, K.M. O'Hara, and J.E. Thomas, Phys. Rev. Lett. \textbf{88}, 120405 (2002)

\bibitem{Takasu03} Y. Takasu, K. Maki, K. Komori, T. Takano, K. Honda, M. Kumakura, T. Yabuzaki, and Y. Takahashi, Phys. Rev. Lett. \textbf{91}, 040404 (2003)

\bibitem{Weber03} T. Weber, J. Herbig, M. Mark, H.-C. N\"agerl, and R. Grimm, Science \textbf{299}, 232 (2003)

\bibitem{Kinoshita05} T. Kinoshita, T.R. Wenger and D.S. Weiss, Phys. Rev. A \textbf{71}, 011602(R) (2005)

\bibitem{Chang06} Ming-Shien Chang, PhD Thesis, http://www.physics.gatech.edu/ultracool/Papers (2006)

\bibitem{OHara01} K.M. O'Hara, M.E. Gehm, S.R. Granade, and J.E. Thomas, Phys. Rev. A, \textbf{64}, 051403(R) (2001)

\bibitem{Hung08} C.-L. Hung, X. Zhang, N. Gemelke and C. Chin, Phys. Rev. A \textbf{78}, 011604(R) (2008)

\bibitem{Brantut08} J.P. Brantut, J.F. Cl\'ement, M. Robert de Saint Vincent, G. Varoquaux, R.A. Nyman, A. Aspect, T. Bourdel, and P. Bouyer, Phys. Rev. A \textbf{78}, 031401(R) (2008).

\bibitem{Petrich94} W. Petrich, M.H. Anderson, J.R. Ensher,and E.A. Cornell, J. Opt. Soc. Am. \textbf{11}, 1332 (1994)

\bibitem{Ketterle93} W. Ketterle, K.B. Davis, M.A. Joffe, A. Martin, and D.E. Pritchard, Phys. Rev. Lett. \textbf{70}, 2253 (1993)

\bibitem{Couvert08} A. Couvert, M. Jeppesen, T. Kawalec, G. Reinaudi, R. Mathevet and D. Guery-Odelin, Europhys. Lett. \textbf{83}, 50001 (2008)

\bibitem{cross_loading}

Actually the tightly confining beam is already on with
$\sim$0.7\,W during the very far-detuned molasses phase. Such a
power corresponds to a light shift similar that created by the
wide beam and the additional trapping volume permits a gain of
$\sim$10\% in the atom number.

\bibitem{power}
An experimental proof of the role of 3 body collisions is the fact
that using more power (14\,W rather than 6\,W) in the confining
beam leads to a lower atom number.

\bibitem{Walraven} O. J. Luiten, M. W. Reynolds, and J. T. M.
Walraven, Phys. Rev. A,\textbf{53}, 381 (1996).

\bibitem{Ketterle96} W. Ketterle and N. J. van Druten, Adv. At. Mol. Opt. Phys. \textbf{37}, 181 (1996)

\bibitem{footnote3body} For background losses, the formula is $\gamma=\eta-4-R\,(\eta-3)$ \cite{Ketterle96}. The difference comes from the fact
that 3-body losses mostly affect the atoms at the trap center
where the density is higher.

\bibitem{1565} To further simplify the degenerated gas apparatus,
all the laser beams for trapping and for cooling could be
generated from a single 1560\,nm fibered diode after amplification
and frequency doubling \cite{Thompson03}.

\bibitem{Thompson03} R. Thompson, M. Tu, D. Aveline, N. Lundblad, and L. Maleki, Optics Express, \textbf{11}, 1709 (2003)

\bibitem{LeCoq06} Y. Le Coq, J.A. Retter, S. Richard, A. Aspect,
P. Bouyer, Appl. Phys. B {\textbf 84}, 627 (2006).
\end{thebibliography}
\end{document}